\documentclass{article}
\usepackage{graphicx}
\usepackage{hyperref}
\usepackage{xcolor}

\usepackage{graphicx} 
\usepackage{hyperref}
\usepackage{xcolor}
\usepackage{setspace}

\usepackage{chngcntr}

\usepackage{geometry}

\newcommand{\REF}[1]{}
\newcommand{\nuevo}[1]{#1}

\makeatletter
\renewcommand{\@maketitle}{%
  \newpage
  \null
  \begin{center}
  \vspace{4mm}
    \let \footnote \thanks
    {\LARGE \@title \par}%
    \vskip 1.5em
    {\large
      \lineskip .5em
      \begin{tabular}[t]{c}
        \@author
      \end{tabular}\par}%
  \end{center}%
  \par
  \vskip 1.5em}
\makeatother

\title{Exploring the interplay of technology, pro-family and prosocial behavior in settlement formation}

\begin{document}
\maketitle

\begin{center}
    {\large    
    Carlos Gracia-L\'azaro$^1$, Alexis R. Hern\'andez$^2$, Felipe Maciel-Cardoso$^1$, Yamir Moreno$^{1,3,4}$\\
    }
    \vspace{3mm}
    $^1$Institute for Biocomputation and Physics of Complex Systems (BIFI), University of Zaragoza, Spain.\\
    \vspace{1mm}
    $^2$ Instituto de F\'{\i}sica, Universidade Federal do Rio de Janeiro, 22452-970 Rio de Janeiro, Brazil.\\
    \vspace{1mm}
    $^3$Department of Theoretical Physics, Faculty of Sciences, University of Zaragoza, Spain.\\
    \vspace{1mm}
    $^4$Centai Institute, Turin, Italy.  
\end{center}

\vspace{1cm}

\begin{abstract}
 \begin{spacing}{2.0} 

  {\large    
    We present an agent-based model that explores the relationship between pro-family and prosocial behaviors and their impact on settlement formation. The objective is to investigate how the technological level and various constraints influence the transition from pro-family to prosocial behavior. The model incorporates factors such as the specialization requirements of the technology, societal tolerance, and dynamic interactions within a synthetic population, where individuals have the choice to prioritize either their family or their own settlement. Agents' fitness is determined by two components: the proportion of pro-family agents within their family and the fraction of prosocial agents in their settlement, as well as its size. Our findings reveal that the transition from pro-family to prosocial behavior is driven by the technological level, and the developmental requirements of the technology shape the smoothness of this transition, ranging from abrupt to gradual. These results emphasize the significance of considering the interplay between the technological level, the nature of the technology, and cultural influences when examining settlement patterns and the dynamics of pro-family and prosocial behaviors in human societies.
 }
 \end{spacing} 

\end{abstract}

\newpage







\section{Introduction}
The settlement patterns arising from the spatial distribution of human populations have long been a subject of study in different fields, such as anthropology, archaeology, human geography, sociology and, more recently, sociophysics \cite{parsons1972archaeological,kowalewski2008regional,ford2009modeling,burley2015bayesian,rodriguez2013modeling,santos2015effect}. Understanding the factors that influence settlement patterns is crucial to gain insights into the dynamics of human societies and their evolution over time. A multitude of factors, including environmental, social, and technological, are known to shape settlement patterns, and their interactions are often complex and nonlinear\cite{maher2012pre}.

The technological level has been recognized as a significant factor influencing settlement patterns \cite{pokotylo1978lithic}. Technological advancements, such as agriculture, transportation, and communication improvements, can enable larger settlements to form by increasing resource availability, facilitating trade and exchange, and enhancing social cooperation \cite{boserup2011conditions,bowles2019neolithic}. However, the relationship between technological level and settlement patterns is not straightforward and can be influenced by various practical constraints, such as resource limitations, space availability, and cultural norms \cite{andrefsky1994raw,smith2004archaeology}. Additionally, social and cultural factors, such as kinship ties, community norms, knowledge sharing, hierarchy, rituals, and tolerance for diversity, can also shape settlement patterns by influencing individual and group behaviors related to family formation, cooperation, and migration \cite{aldenderfer1993ritual,boyd2011cultural,dandekar2011satvai,henderson2005muisca}.

Cooperation plays a crucial role in group formation and has been widely studied in the context of human societies \cite{tanimoto2007does,pan2013cooperation,cuesta2015reputation,javarone2017evolutionary,perc2017statistical,javarone2016statistical,szekely2021evidence}. In this sense, public goods games, a common theoretical and experimental paradigm used to study cooperation, have provided insights into the mechanisms underlying cooperative behavior in groups \cite{szabo2002phase,hauert2008public,szolnoki2010reward}, and also through family or kin relationships \cite{molina2019intergenerational}. In public goods games, individuals can choose to cooperate by contributing to a common pool of resources that benefits the group as a whole, or they can choose to free-ride and not contribute while still benefiting from the contributions of others \cite{perc2013evolutionary}. 

Altruistic behaviors, such as cooperation and helping others without expecting anything in return, have been widely studied in the context of human societies and can enhance social cohesion and cooperation within groups, leading to the formation of larger and more stable settlements \cite{bowles2006group}. To better understand the complex interplay between the technological level and cultural factors in shaping settlement patterns, agent-based modeling (ABM) provides a powerful approach. ABM is a computational modeling technique that allows for the simulation of individual-level behaviors and interactions within a larger social system \cite{axelrod2006agent,epstein2006remarks}. ABM has been widely used to study settlement patterns and human societies, as it provides a flexible and dynamic framework to investigate the emergent properties of social systems \cite{schelling1971dynamic,axelrod1997dissemination,parker2003multi}, as well as the socio-economical evolution of societies \cite{gracia2022dynamics} and resources sharing \cite{santos2015effect}. By representing individuals as agents with decision-making roles and simulating their interactions over time, ABM can capture the nonlinear and evolving nature of settlement dynamics and provide insights into the mechanisms underlying observed patterns.

In this paper, we present an agent-based model to study the relationship between pro-family and prosocial behaviors and their influence on settlement formation. Pro-family behavior refers to actions that prioritize the well-being and survival of one's family, such as caring for one's spouse, children, and close relatives. Prosocial behavior, on the other hand, encompasses actions that benefit the larger society or community, such as cooperation, sharing, and helping others. Both pro-family and prosocial behaviors are fundamental to human societies and play critical roles in shaping social dynamics and settlement patterns\cite{johnson2000evolution,bowles2006group,bowles2019neolithic}.

We focus on understanding how the transition from pro-family to prosocial behavior is influenced by the technological level and other constraints, such as the development requirements (i.e., the population size required for a given technology to achieve optimal performance) and cultural issues. Technological advancements have been recognized as a significant driver of societal change, affecting settlement patterns, resource utilization, and social behaviors\cite{ellis2015ecology}. However, the relationship between technological level and settlement patterns is complex and multifaceted, with cultural factors also influencing settlement dynamics. For example, higher technological levels may enable the formation of larger settlements through increased resource extraction and production capacities, but resource and space limitations may hinder settlement growth beyond a certain point. Additionally, cultural norms, beliefs, and values may impact individuals' decisions on whether to prioritize their families or the larger society, shaping settlement patterns in the process.

To investigate these dynamics, we develop an agent-based model that simulates a synthetic population of individuals who can choose between promoting their family or settlement. The fitness of agents is determined by two components: one based on the fraction of pro-family agents in their family and the other based on the fraction of prosocial agents in their settlement and its size. We incorporate technological level, development requirements, and cultural factors as parameters in the model to explore their effects on settlement patterns and the transition from pro-family to prosocial behavior.

The contributions of this paper are twofold. First, we develop an agent-based model that integrates pro-family and prosocial behaviors, technological level, and other constraints to investigate settlement dynamics. This model provides a novel approach to studying the interplay between individual behaviors, technological advancements, and practical limitations in shaping settlement patterns. Second, we present findings from our simulations that shed light on the complex relationship between technological level and settlement patterns, highlighting the importance of considering specialization requirements, cultural, and technological factors together in understanding settlement dynamics in human societies. \nuevo{Our main result is that the technological level drives the transition from pro-family to prosocial behavior. Furthermore, for high values of tolerance toward non-prosocial behavior, the transition changes from abrupt to gradual as the population demands of the technology increase. On the other hand, for intolerant societies towards non-prosocial behavior, the transition is abrupt regardless of the technology demands.}

The rest of the paper is organized as follows. In Section \ref{TheModel}, we describe the details of our agent-based model, including the assumptions, parameters, and rules governing agent behavior. We aim to provide a comprehensive understanding of the model's components to establish a solid foundation for our subsequent analyses. In Section \ref{Results}, we present analytical considerations alongside the results of our simulations. By combining these two approaches, we can give a detailed assessment of the outcomes and implications of our study. Finally, in Section \ref{Conclusions}, we conclude with a summary of our findings and their implications, and we also offer prospective remarks on potential avenues for further research. 

\section{Modeling the emergence of societal clusters}
\label{TheModel}



We propose an agent-based model to study the tension between pro-family and prosocial behaviors and its effect on settlement formation. To this end, we consider a synthetic population of $n$ agents grouped in families, which in turn may aggregate in settlements. Agents can adopt two possible strategies: foster either family or settlement. In this way, the model captures the trade-off between pro-family and prosocial behaviors by allowing agents to choose between promoting their family or their settlement.

Let us denote the families by subscripts and the settlements by superscripts. The fitness of an agent, denoted by $\pi_f^s$, is determined by two components. The first component is a function of the fraction of pro-family agents in its family $f$. The second component is a function of the fraction of prosocial agents in its settlement $s$, as well as the population of the settlement. Specifically, the fitness is given by:

\begin{equation}
\pi_f^s=(1-\rho_f) + \alpha \frac{n^s}{1+n^s/h} \rho^s\;,
\label{eq_fitness}
\end{equation}
where $\rho_f$ is the fraction of prosocial agents in family $f$,  $\rho^s$  the fraction of prosocial agents in settlement $s$ and $n^s$ the number of agents in settlement $s$. As model parameters, $\alpha$ represents the efficiency of societies (the technological level), and $h$ corresponds to a saturation term that takes into account the nature of the technology, as explained in the next paragraph.

According to the last term of Eq. \ref{eq_fitness}, the fitness associated with prosocial behavior increases with the size of the population in the settlement. This increase captures the wealth obtained by the community through task distribution, specialization, and collective work. However, this growth is not unlimited and has an asymptotic limit represented by the saturation term ($h$), which is determined by the nature of the technology. A low value of $h$ indicates that once a settlement has reached a relatively small size, the needs for specialization have been met, and further settlement growth will not significantly increase the relative benefits. Conversely, a high value of $h$ corresponds to a technology that allows for greater levels of specialization, enabling larger settlements to reap more benefits. While the saturation term $h$ is inspired by the idea of carrying capacity \cite{verhulst1845resherches,del2004carrying}, in this context, it represents the population size required for a given technology to achieve optimal performance; besides, the technological level $\alpha$ is related to the degree of implementation of the technology.



It is important to note that all the agents within a given family equally perform, i.e., the model presumes that, within a family, the resources and wealth are shared (\textit{in a family, everyone eats from the same pot}). \nuevo{Furthermore, the coefficient $\alpha$ plays a crucial role in determining the efficiency of societies, reflecting the level of technological advancement. This coefficient is not arbitrarily set but emerges from the reduction of two underlying coefficients: those associated with the family-driven and settlement-driven fitness components. The family-driven fitness component (first term on the right-hand side of Equation \ref{eq_fitness}) represents the advantage that an agent gains within its family based on its strategy and the strategies of other family members. Similarly, the settlement-driven fitness component (last term of Equation \ref{eq_fitness}) captures the advantage an agent experiences within its settlement due to the prevalence of prosocial behavior in that settlement. Both components contribute to the overall fitness of the agent, influencing its probability of reproducing and passing on its strategy to the next generation.}

\nuevo{In the model, without loss of generality under the bellow defined dynamics, we have simplified these two underlying coefficients to a single one: $\alpha$. Specifically, we set the family-driven fitness coefficient to one, condensing the weights of the family-driven and settlement-driven fitness components into $\alpha$. As $\alpha$ increases, settlements with a higher proportion of prosocial agents and larger populations gain an advantage in the reproduction process.}






\nuevo{Regarding the dynamics of the model, it involves two fundamental processes: death and reproduction. At each time step, a random individual is chosen for death from the entire population with an equal probability for each agent. Following the death event, a new individual is selected for reproduction among all the agents, with the probability of selection being proportional to the individual's fitness. This reproduction process follows the Moran rule, which means that fitter individuals have a higher chance of reproducing and passing on their strategy to the next generation.}

\nuevo{Additionally, we introduce two critical mechanisms that impact settlement formation and dynamics. The first mechanism addresses the family size limitation. If a family grows beyond a maximum size threshold denoted by $\theta$, it splits into two new families, both of which remain part of the same settlement. This splitting process allows families to maintain a manageable size, adhering to realistic family size limitations.}

\nuevo{The second mechanism involves the viability of settlements based on the prevalence of prosocial behavior within them. If in a settlement the density of prosocial agents is lower than a threshold $\xi$, the settlement becomes unfeasible and collapses. When a settlement collapses, each of its families becomes a new settlement on its own. The parameter $T=1-\xi$ represents the settlements' tolerance to non-prosocial behavior. Higher values of tolerance $T$ involve that settlements are more resilient to the presence of non-prosocial agents, while lower values of $T$ indicate that settlements are more sensitive to deviations from full prosociality.}

\nuevo{By incorporating these elements, the model simulates the evolution of settlements over time, capturing the interplay between individual fitness, family dynamics and settlement formation to shed light on how technological advancement and population requirements influence settlement patterns and the prevalence of prosocial behavior in human societies.}

\section{Results and discusion}
\label{Results}

\subsection*{Analytical considerations}

Before delving into the numerical results, let us explore some minimal conditions for prosociality to thrive. From Eq. \ref{eq_fitness}, we can deduce the total fitness $\Pi^s$ of a settlement $s$ of size $n^s$ and prosocial density $\rho^s$:
\begin{equation}
\Pi^s=\sum_{i\in s} \left[ (1-\rho_{f(i)}) + \alpha \frac{n^s}{1+n^s/h} \rho^s\right] \;,
\label{eq_settlement_fitness}
\end{equation}
where index $i$ runs over all the agents in settlement $s$, and $\rho_{f(i)}$ refers to the density of prosocial agents in the $i$'s family. After straightforward manipulation, we obtain the equivalent expression: 

\begin{equation}
\Pi^s /n^s = 1+\left(\alpha \frac{n^s}{1+n^s/h}-1\right)\rho^s \;.
\label{eq_settlement}
\end{equation}

For agents in the settlement to have greater mean fitness than those in other settlements with a lower fraction of prosocial agents, the term in parentheses on the right side of Equation \ref{eq_settlement} has to be positive. This provides a criterion for the selective pressure toward the settlements with a higher prosocial ratio:

\begin{equation}
\alpha \frac{n^s}{1+n^s/h} -1 > 0 \;.
\label{eq_minimum_settlement_size}
\end{equation}

From Equation \ref{eq_minimum_settlement_size}, we obtain the expression:

\begin{equation}
n^s > \frac{1}{\alpha-1/h} \;,
\label{eq_minimum_settlement_size2}
\end{equation}
which constitutes the minimum settlement size required for prosocial behavior to be advantageous. Nevertheless, within each settlement, families with a higher proportion of pro-family members will have greater fitness compared to families with a lower proportion. Thus, provided there are settlements that meet the condition outlined in Equation \ref{eq_minimum_settlement_size2}, there will be a trade-off between i) the selective pressure towards prosocial behavior because of the fitness differences among settlements and ii) the advantage of pro-family behavior within settlements. Conversely, if no settlement satisfies the condition, meaning they are below the minimum required size, all the selective pressure will be directed toward pro-family behavior. In addition to those mechanisms, the dynamic within families exhibits a short temporal scale: due to the limited size of families, fluctuations inside each family quickly result in an absorbing configuration where all members adopt the same strategy.


There is another threshold value beyond which, within a family, an agent will earn more by being prosocial than by being pro-family. To explain this, let's consider an agent $i$ who has not yet decided on its strategy. Let $n_{f(i)}$ be the $i$'s family size, $x$ the number of pro-family members in $i$'s family, $n^{s(i)}$ the $i$'s settlement size, and $y$ the number of prosocial agents in $i$'s settlement. Before considering $i$'s strategy, agent $i$ has as provisional fitness, $\pi^*(i)=x/n_{f(i)}+\mu y/n^{s(i)}$, where $\mu=\alpha n^{s(i)} / (1+n^{s(i)}/h)$. 

If agent $i$ decides to be pro-family, its fitness will be $\pi(i)=\pi^*(i)+1/n_{f(i)}$; conversely, if agent $i$ decides to be pro-social, its fitness will be $\pi(i)=\pi^*(i)+\mu/n^{s(i)}$.  For agent $i$ to have an advantage in being pro-social, the condition $\mu/n^{s(i)} > 1/n_{f(i)}$,  must be satisfied, from which, by recovering the expression for $\mu$, we obtain:

\begin{equation}
n_{f(i)} > \frac{1+n^{s(i)}/h}{\alpha} \;.
\label{eq:condition_best_strategy}
\end{equation}

This means that, in small settlements, there is a critical size for the family size beyond which prosocial behavior becomes more advantageous than pro-family behavior within the settlement; this critical value depends on the family and settlement sizes. However, it is essential to note that the condition given by Equation \ref{eq:condition_best_strategy} does not eliminate the selective pressure toward pro-family behavior within a settlement because the proposed dynamics don't allow agents to change their minds (nonetheless, it is still a condition worth considering, as it may have implications for other dynamics within the system). Within a given settlement, more successful agents, who are those in pro-family-prone families, are always more likely to have offspring than those prosocial-prone, although the difference in fitness will eventually become negligible for large enough settlements.

\subsection*{Numerical results}

The initial conditions for all the simulations presented in the main text were equiprobable, with 50\%-50\% of agents adopting prosocial and pro-family behaviors, respectively. However, to check the robustness of the results, in Section II of the Supplementary Information, we provide additional simulations with different initial conditions, specifically with only 5\% (resp., 95\%) of agents adopting prosocial (resp., pro-family) behavior. Also, this 5\%-95\% initial condition will allow us to explore the cases where prosocial behavior is initially residual, mimicking realistic scenarios where families form settlements only with favorable conditions. Additionally, while the simulations presented in the main text were conducted with a population size of 10,000 agents, we also examined larger system sizes in Section I of the Supplementary Information to further validate the robustness of our findings. Lastly, we take $\theta=10$, which provides a maximum family size of 10 members and characteristic size of $\sim5$ members.

\begin{figure}
  \centering
    \includegraphics[width=0.32\textwidth]{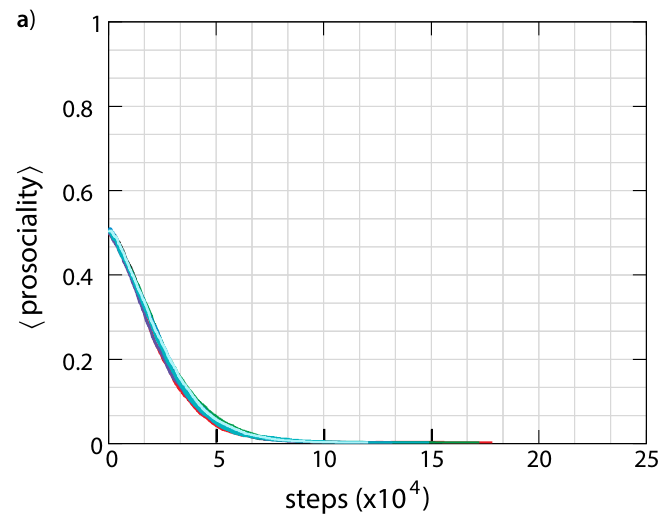}
    \includegraphics[width=0.32\textwidth]{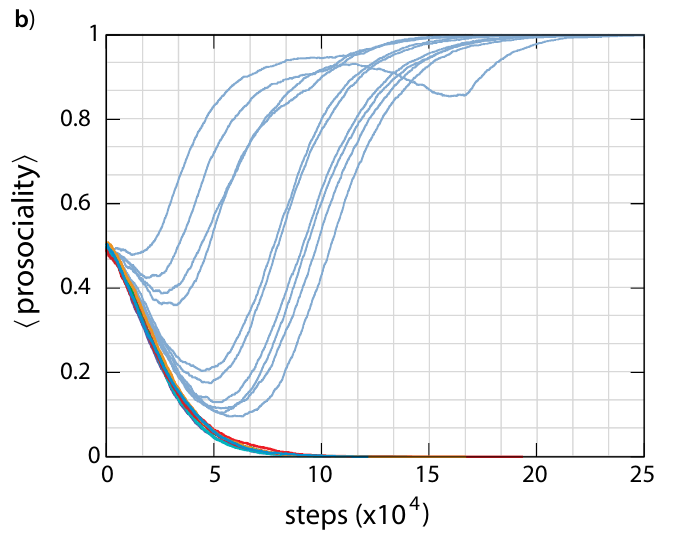}
    \includegraphics[width=0.32\textwidth]{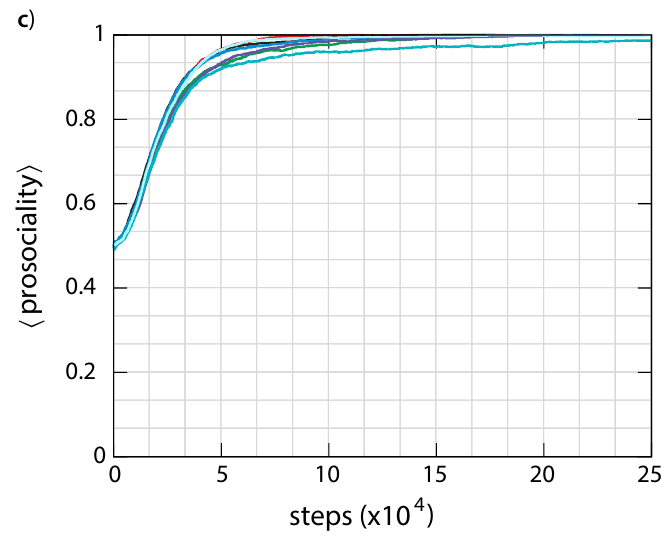}
  \caption{Time evolution of the fraction of prosocial agents for three different technological levels: $\alpha=0.065,0.085, 0.5$ in panels \textbf{a}, \textbf{b}, and \textbf{c}, respectively. Each curve represents an independent simulation. In Panel \textbf{a} ($\alpha=0.065$), all simulations converge to a pro-family state, while in Panel \textbf{b} ($\alpha=0.085$), bistability is observed, with some simulations leading to a prosocial state and others to a pro-family one. In Panel \textbf{c} ($\alpha=0.5$), all simulations converge to a prosocial state. The saturation term was fixed to $h=20$, tolerance to $T=0.5$, and the population size to $n=10^4$.}
  \label{fig:time_evolution}
\end{figure}

We begin our analysis of the numerical results by examining the time evolution of the fraction of prosocial agents in different scenarios, as illustrated in Figure \ref{fig:time_evolution}. Each panel in Figure 1 represents a different technological level $\alpha$, showing the effect of $\alpha$ on the prevalence of prosocial behavior over time, with representative independent simulations depicted as curves: the x-axis represents the time (step) and the y-axis the ratio of prosocial agents. As shown in Panel \textbf{a} ($\alpha$=0.065), at a low technological level, pro-family behavior outperforms prosocial behavior, resulting in a dominant prevalence of pro-family agents throughout the simulations. This observation suggests that at low levels of technological advancement, prosocial behavior may not offer sufficient advantages in terms of societal performance, and families with a high pro-family orientation may have a competitive advantage.

Moving on to Panel \textbf{b} of Figure \ref{fig:time_evolution} ($\alpha$=0.085), it is observed that pro-family behavior dominates during the initial stages of the simulation, as indicated by the progressive prevalence of pro-family agents compared to prosocial ones. However, as the simulation progresses, a new phenomenon emerges: in some simulations, after reaching a minimum, the level of prosociality begins to increase. Due to the splitting of families and the formation of new settlements, some of these newly formed settlements exhibit a higher number of prosocial agents compared to others with a lower prosocial level. For certain instances, these prosocial-proned settlements overcome the critical size. It is important to note that, within these settlements with high prosociality, families with a high ratio of pro-family agents still perform better than prosocial families, as in any settlement. However, at the same time, regarding the balance between settlements, the settlements with high prosociality overcome those with high pro-family behavior, suggesting a possibility of a dynamic shift from pro-family to prosocial behavior in the system. This observation implies that the dynamics of the system are complex and dependent on the interplay between technological level, agent behavior, and societal performance. The settlements with a higher proportion of prosocial agents tend to grow in size, as the advantages of prosocial behavior become apparent in terms of societal performance. This leads to an increase in the prevalence of prosocial behavior over time, ultimately resulting in an evolution toward a prosocial society. This intriguing result underscores the importance of considering the dynamic nature of the system, specifically the comparison between different settlements, in understanding the emergence and evolution of prosocial behavior.

Finally, in Panel \textbf{c} of Figure \ref{fig:time_evolution}, where the technological level is high ($\alpha$=0.5), a different tendency is observed, with prosocial behavior dominating throughout the simulations. The trend is monotonous, with a consistently higher prevalence of prosocial agents compared to pro-family agents over time. This result suggests that for high values of technological advancement, prosocial behavior provides significant advantages in terms of societal performance, leading to its prevalence in the system. The consistent dominance of prosocial behavior in Panel \textbf{c} further emphasizes the potential positive influence of the technological level on shaping prosocial behavior in society. Overall, this Figure \ref{fig:time_evolution} provides valuable insight into the relationship between technological advancement and prosocial behavior over time, showing a transition from pro-family to prosocial behavior as the technological level $\alpha$ increases.

\begin{figure}
  \centering
    \includegraphics[width=0.48\textwidth]{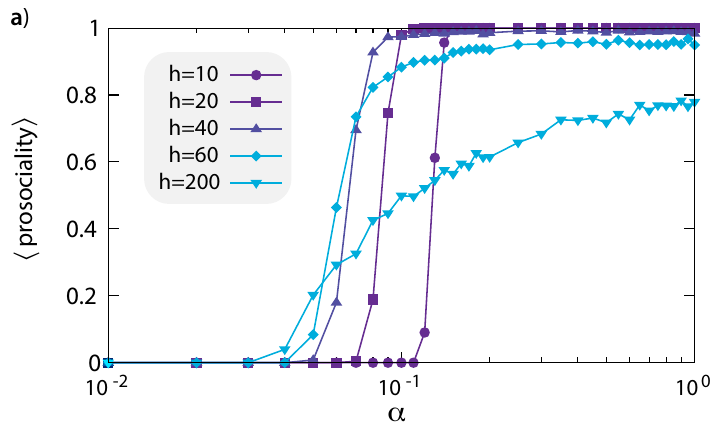}
    \includegraphics[width=0.48\textwidth]{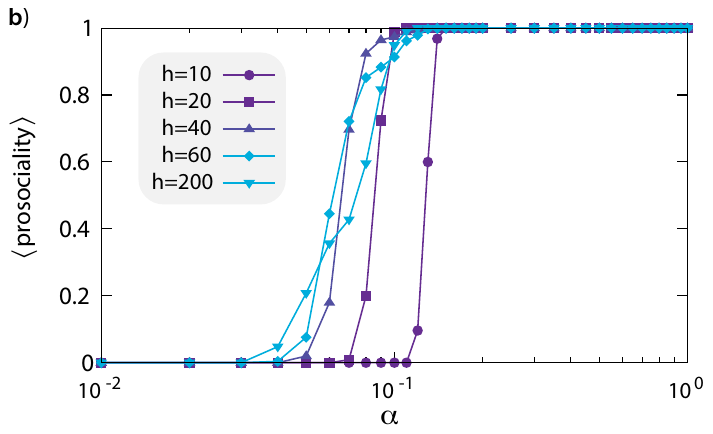}
    \includegraphics[width=0.48\textwidth]{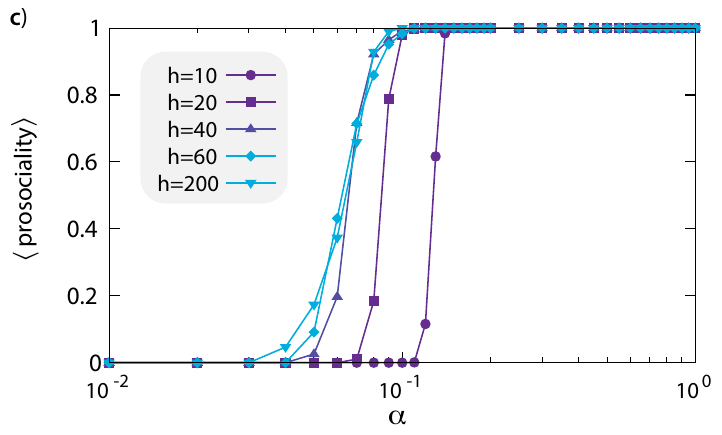}
    \includegraphics[width=0.48\textwidth]{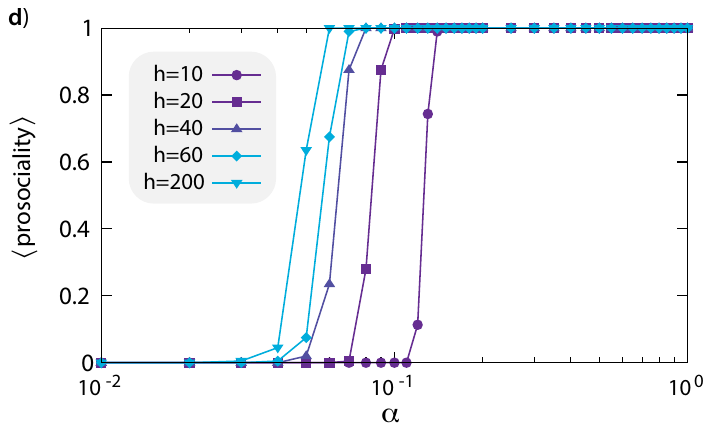}
  \caption{Final fraction of prosocial agents as a function of the technological level $\alpha$. Panels \textbf{a}, \textbf{b}, \textbf{c}, and \textbf{d} display the results for a tolerance of $T=1,0.95,0.9, 0.5$, respectively. Each series corresponds to a different saturation value $h=10, 20, 40, 60, 80, 200$, and each point represents the mean value over 1000 numerical simulations. $n=10^4$. See the text for further details.}
  \label{fig:c_Vs_alpha}
\end{figure}

In order to study the transition from pro-family to prosocial behavior in more detail, we analyzed the state of the system after the transient period for different values of the technological level $\alpha$. Figure \ref{fig:c_Vs_alpha} presents the stationary mean value, averaged over 1000 simulations, of the ratio of prosocial agents as a function of $\alpha$, with each panel corresponding to a different tolerance $T$. Consistent with previous findings, it is observed that prosocial behavior tends to increase with higher levels of technological advancement.

In Panel \textbf{a} of Figure \ref{fig:c_Vs_alpha} ($T=1$), it is shown that as the saturation term $h$ increases, the curve of $\langle$prosociality$\rangle$ versus $\alpha$ exhibits a smoother transition for high values of tolerance, that is, in the absence of collapse events ($T=1$), a higher saturation term allows for a more gradual shift from pro-family to prosocial behavior as the technological level increases.

In subsequent panels (\textbf{b}, \textbf{c}, and \textbf{d}) of Figure \ref{fig:c_Vs_alpha}, where the tolerance decreases (i.e., $T=0.95, 0.9,$ and $0.5$ respectively), a distinct trend becomes evident. As tolerance decreases, the transition from pro-family to prosocial behavior becomes sharper for higher saturation terms ($h$). In Panels \textbf{b} and \textbf{c}, it is observed that the transition becomes more abrupt for $h>40$, with a steeper increase in prosocial behavior as $\alpha$ increases. 
Finally, in Panel \textbf{d} of Figure \ref{fig:c_Vs_alpha} ($T=0.5$), the transitions are abrupt for any saturation term $h$. This result indicates that, at low enough tolerance levels, where settlements need to exhibit a significant prosociality to avoid collapse, even a small increase in the technological level $\alpha$ can trigger a rapid shift from pro-family to prosocial behavior in the system.

These findings highlight the complex interplay between technological level, tolerance, saturation term, and the transition from pro-family to prosocial behavior. Higher saturation terms and tolerance levels result in a smoother transition, while lower tolerance levels intensify the transition dynamics and make the transitions more abrupt. This suggests that the relationship between technological advancement and the transition to prosocial behavior is influenced by tolerance and saturation terms.

\begin{figure}
  \centering
    \includegraphics[width=0.48\textwidth]{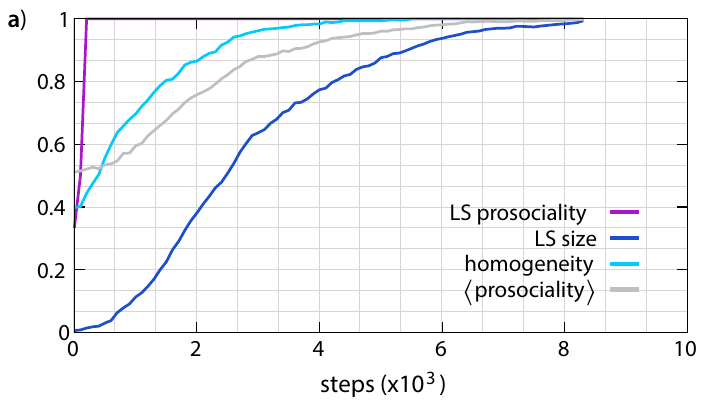}
    \includegraphics[width=0.48\textwidth]{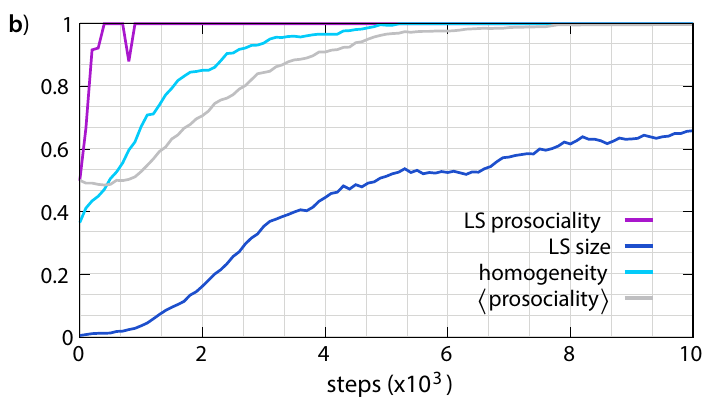}
    \includegraphics[width=0.48\textwidth]{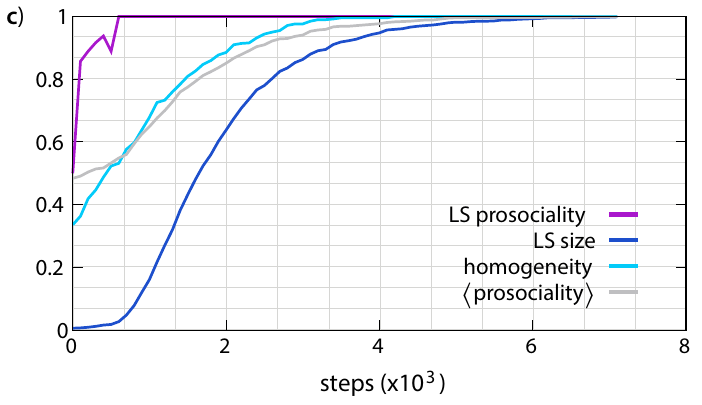}
    \includegraphics[width=0.48\textwidth]{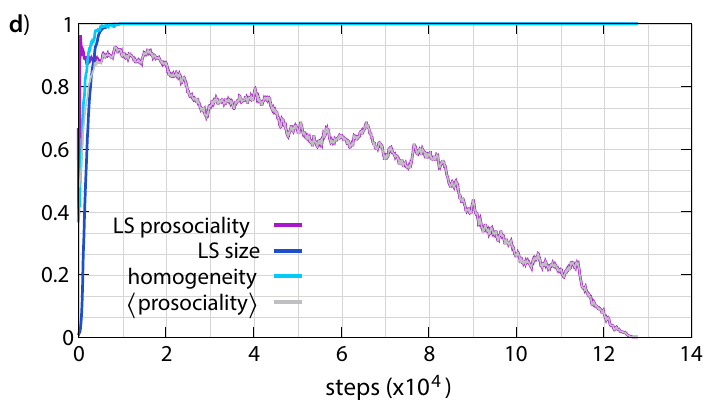}
    \includegraphics[width=0.48\textwidth]{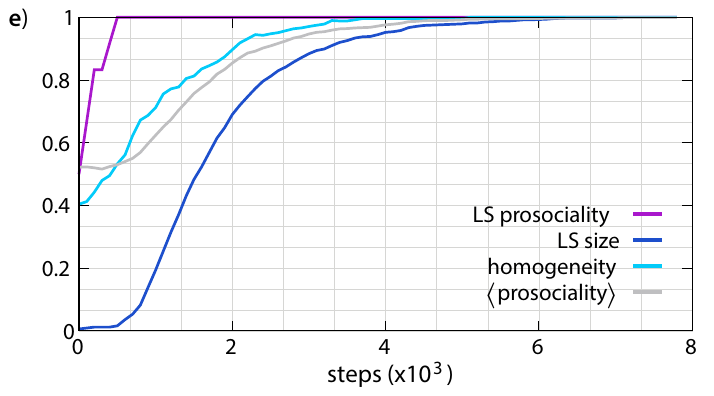}
    \includegraphics[width=0.48\textwidth]{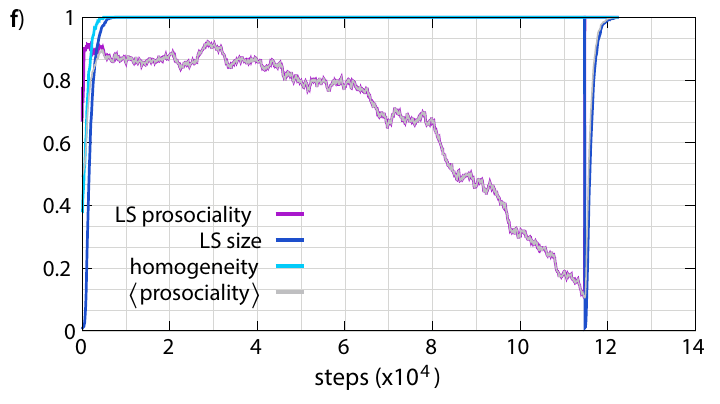}
    \caption{Time evolution for a technological level beyond its critical value ($\alpha=0.4$). Two independent realizations are shown for each choice of the parameters. In each row, the left and right panels represent two different characteristic simulations for the same parameters, one per panel. Panels \textbf{a}-\textbf{b} correspond to a tolerance of $T=1$, $h=20$; panels \textbf{c}-\textbf{d} to $T=1$, $h=200$; and panels \textbf{e}-\textbf{f} to $T=0.9$, $h=200$. Lines represent, respectively, the fraction of prosocial agents in the largest settlement (LS prosociality), the size of the largest settlement (LS size), family homogeneity in the entire system (homogeneity), and the fraction of prosocial agents in the system ($\langle$prosociality$\rangle$). The x-axis denotes the time steps in the simulation. Population size:  $n=10^4$. See the text for further details.
    }
  \label{fig:transient_detail}
\end{figure}

To provide a clearer explanation of the previous results, let us revisit the transient dynamics to gain a deeper understanding of the different types of transitions observed. Figure \ref{fig:transient_detail} illustrates different transient behaviors for a technological level beyond its critical value ($\alpha=0.4$). Each row in the figure represents two characteristic simulations with identical parameters, one simulation per panel. Panels \textbf{a}-\textbf{b} correspond to a tolerance of $T=1$ and a saturation term of $h=20$, panels \textbf{c}-\textbf{d} to $T=1$ and $h=200$, and panels \textbf{e}-\textbf{f} to $T=0.9$ and $h=200$. \textit{LS prosociality} line in each panel represents the fraction of prosocial agents in the largest settlement, shedding light on the prevalence of prosocial behavior. \textit{LS size} line depicts the size of the largest settlement. \textit{Prosociality} line represents the fraction of prosocial agents in the entire system, providing an overview of the overall distribution of prosocial behavior. Additionally, $\langle$\textit{homogeneity}$\rangle$ line represents the family homogeneity $U$ in the entire system, which is defined as:

\begin{equation}
U= \frac{2}{F}\sum_j \left| 0.5-\rho_{f(j)} \right|,
\label{eq_H}
\end{equation}
where the index $j$ runs over all families, and $F$ is the total number of families in the society. Family homogeneity $U$ measures the degree of uniformity among agents within each family. If all agents in a given family share the same strategy, that family is considered homogeneous (and will contribute toward $U=1$); conversely, if half of the members of a family have a strategy and the rest have the opposite, that family has maximum heterogeneity and will contribute toward $U=0$.

In panels \textbf{a} and \textbf{b} of Figure \ref{fig:transient_detail}, corresponding to $h=20$, the system consistently tends toward prosociality. However, two different behaviors are observed. In Panel \textbf{a}, the largest settlement rapidly achieves complete prosociality in its early stages, growing to dominate the entire society. In other realizations (as that one displayed in Panel \textbf{b}), the more prosocial settlements take slightly longer to reach complete prosociality, allowing for the growth of different relatively prosocial settlements that alternate in size leadership (indicated by the valleys in the \textit{LS prosociality} curve). This slower growth enables the coexistence of multiple prosocial settlements. Although the simulations shown are for $T=1$, similar behaviors are observed for $T<1$ (not shown here for brevity), as tolerance does not play a decisive role in settlements with significant prosociality.

However, for high saturation term values (panels \textbf{c} through \textbf{f}, $h=200$), tolerance becomes a key factor. Panels \textbf{c}-\textbf{d} depict two characteristic simulations for $T=1$ (no collapses). It is important to note that in the case of high $h$ values, reaching the optimal level of developmental performance relies on significantly large settlements. In some realizations, when the largest settlement gets the maximum size, it is already fully prosocial (Panel \textbf{c}). In other cases, the largest settlement reaches the maximum size without being fully prosocial (Panel \textbf{d}). In this latter scenario, the only selective pressure present is intra-settlement competition, which always favors pro-family behavior since it yields higher fitness within the single settlement. Without collapses, prosociality will decline in this single settlement until it reaches zero. On the other hand, when settlements are allowed to collapse ($T<1$, panels \textbf{e}-\textbf{f}), both of the previously mentioned scenarios will ultimately lead to a prosocial society: similar to before, in some realizations, a settlement reaches the maximum size being already fully prosocial (Panel \textbf{e}), while in the cases it reaches the maximum size without being fully prosocial (Panel \textbf{f}), prosociality will decline until a value of $1-T$, at which point the settlement will collapse. The collapse is depicted by a sharp decrease in the curve corresponding to the size of the largest settlement, shifting from 1 to almost zero. At the moment of the collapse, as families are already mono-strategic ($U=1$), there will be some completely prosocial families. These prosocial families have a significant evolutionary advantage over exclusively pro-familiar settlements, as there are no mixed settlements. Therefore, after the collapse, some of them will establish the first prosocial multifamily settlements, as indicated by a sharp increase in the curve representing the prosociality of the largest settlement. This substantial difference in fitness causes the largest prosocial settlement to grow and dominate the system rapidly.


As a corollary, provided a high enough technological implementation degree, a minimum level of intolerance towards non-prosocial behavior is enough for prosocial behavior to dominate society irrespective of the technological specialization requirements.

\begin{figure}
  \centering
    \includegraphics[width=0.48\textwidth]{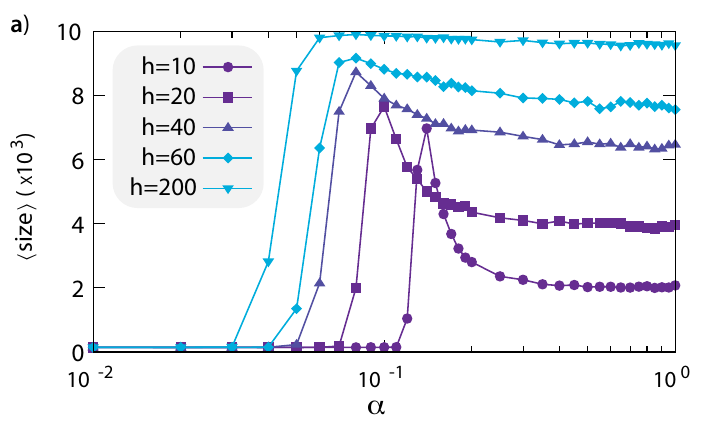}
    \includegraphics[width=0.48\textwidth]{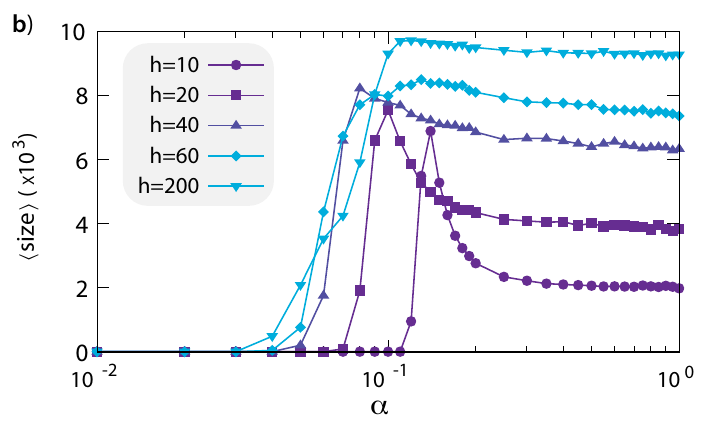}
    \includegraphics[width=0.48\textwidth]{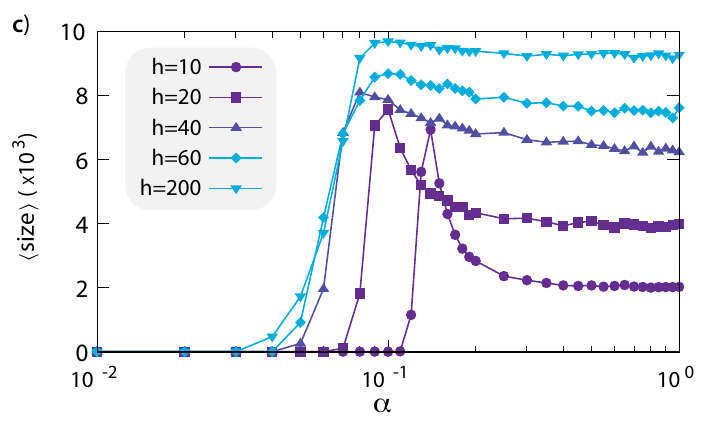}
    \includegraphics[width=0.48\textwidth]{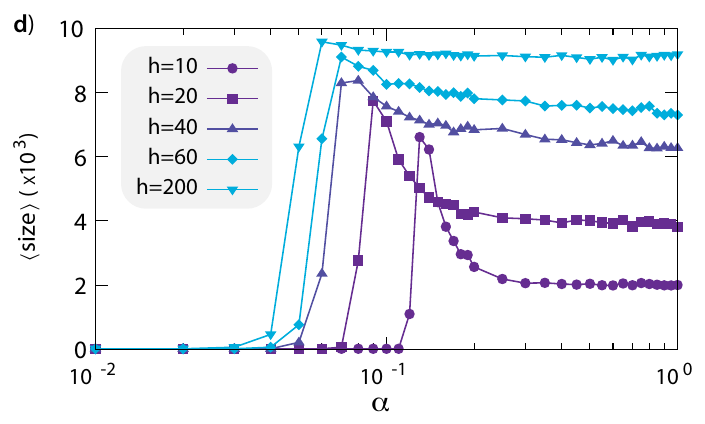}
  \caption{Size of the largest settlement as a function of the technological level $\alpha$. Panels \textbf{a}, \textbf{b}, \textbf{c}, and \textbf{d} display the results for a tolerance of $T=1,0.95,0.9, 0.5$, respectively. Each point represents the mean value over 1000 numerical simulations. $n=10^4$.}
  \label{fig:largest_settlement_Vs_alpha}
\end{figure}

Regarding the aggregation of the population in settlements, Figure \ref{fig:largest_settlement_Vs_alpha} presents the size of the largest settlement in relation to the technological level for tolerance values of $T=1, 0.95, 0.9,$ and $0.5$ in panels \textbf{a}, \textbf{b}, \textbf{c}, and \textbf{d}, respectively. As shown, low values of the technological level $\alpha$ involve a scattered population. In this disaggregated phase corresponding to low $\alpha$, the size of the largest settlement in the absence of collapses ($T=1$, panel \textbf{a}) is greater (approximately 21 agents and 5 families) than in scenarios where collapses are possible ($T<1$, panels \textbf{b}-\textbf{d}) where for technology levels below the critical value, settlements are typically single-family, with an average of approximately 4 members. These findings are consistent with the fact that collapse events limit the growth of pro-family settlements.

Beyond a critical value, the largest settlement size increases, although the curves exhibit non-monotonic behavior, with a local maximum observed for intermediate technological levels. The peak becomes more pronounced as the saturation term $h$ decreases. The observed decrease in the size of the largest settlement as $\alpha$ surpasses its peak, especially for low values of $h$, agrees with the transient previously analyzed, highlighting that while a high technological level can facilitate the formation of large societies, the time scales associated to intra-family, intra-settlement and global dynamics play a key role in the coexistence of several prosocial settlements. It should be noted that the saturation term $h$, along with an asymptotic limit for the enhancement by settlement growth, determines a practical value beyond which an increase in size does not confer a significant advantage. As a result, in the presence of such limitations, settlements tend to be medium-sized instead of large, resulting in a settlement distribution pattern that reflects a balance between technological progress and practical constraints. For technologies with moderate development requirements, settlements may remain relatively small and dispersed, even with high levels of implantation of the technology.


\begin{figure}
  \centering
    \includegraphics[width=0.48\textwidth]{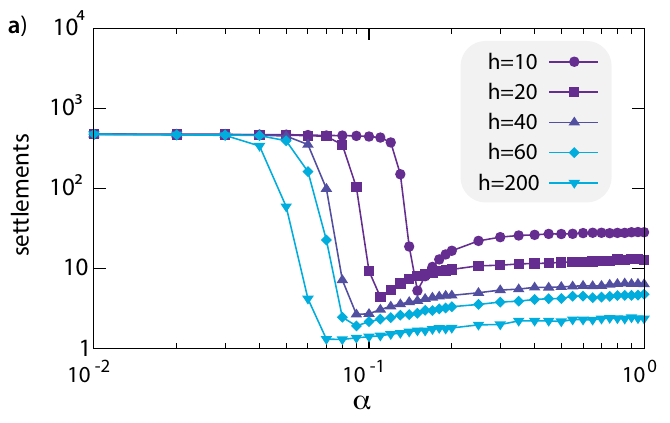}
    \includegraphics[width=0.48\textwidth]{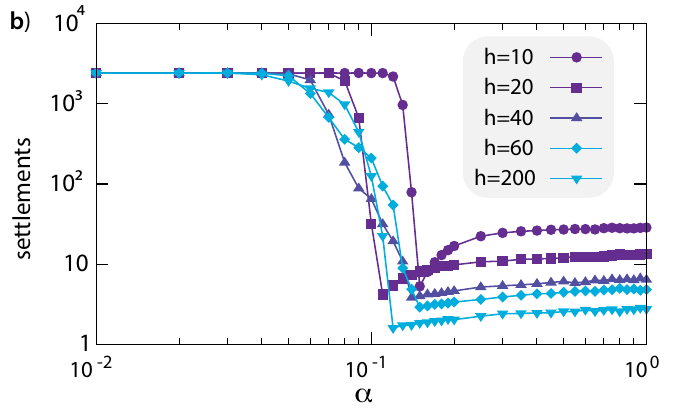}
    \includegraphics[width=0.48\textwidth]{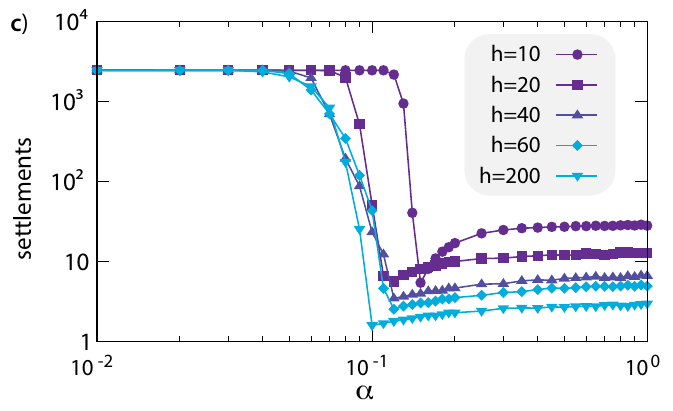}
    \includegraphics[width=0.48\textwidth]{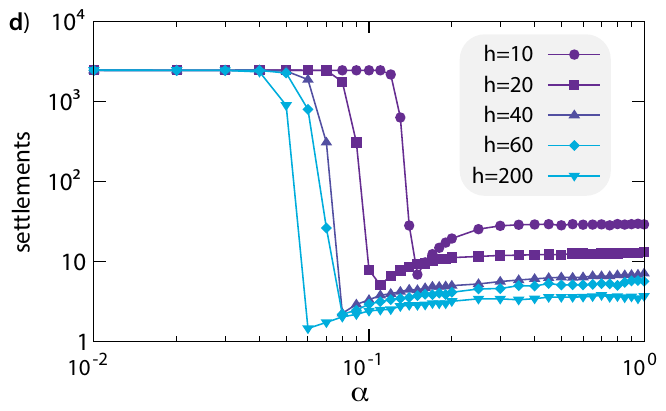}
  \caption{Number of settlements versus the technological level $\alpha$. Panels \textbf{a}, \textbf{b}, \textbf{c}, and \textbf{d} display the results for a tolerance of $T=1,0.95,0.9, 0.5$, respectively. Each series corresponds to a different saturation value $h=10, 20, 40, 60, 80, 200$, and each point represents the average over 1000 numerical simulations. $n=10^4$.}
  \label{fig:settlements_Vs_alpha}
\end{figure}

Figure \ref{fig:settlements_Vs_alpha} shows the number of settlements as a function of technological level, with each panel corresponding to a specific value of tolerance ($T=1, 0.95, 0.9,$ and $0.5,$ respectively). For low values of technological level $\alpha$, the number of settlements remains high, corresponding to a scattered population. However, as $\alpha$ increases, the number of settlements decreases, although the trend is not monotonic. The plot shows a minimum in the number of settlements for intermediate technological levels, which represents the point where the population is most concentrated and occurs at the value of $\alpha$ that depicts a maximum settlement size. In accordance with the results shown in the previous figure, the minimum value becomes lower as the saturation term $h$ increases, while the difference between the minimum and surrounding values for higher $\alpha$ becomes more pronounced as $h$ decreases. Furthermore, the scenario without collapses ($T=1$, panel \textbf{a}) exhibits a lower number of settlements for low technological levels $\alpha$ compared to the cases where collapses may occur ($T<1$, panels \textbf{b}-\textbf{d}). These differences in the scattering phase (low $\alpha$) between $T=1$ and $T<1$ are consistent with the previously discussed observation that in high-tolerance scenarios ($T=1$), settlements in the disaggregated phase have an average of approximately five families, whereas, in lower tolerance scenarios ($T<1$), settlements tend to consist of single families due to collapse events limiting the growth of settlements.

The comparison of figures \ref{fig:largest_settlement_Vs_alpha} and \ref{fig:settlements_Vs_alpha} provides insight into the relationship between the size of the largest settlement and the number of settlements. The graphs show that as the technological level $\alpha$ increases, the number of settlements initially decreases but eventually reaches a local minimum. At the same time, the size of the largest settlement increases and also reaches a local minimum for intermediate values of $\alpha$. These patterns suggest that technology with low development requirements can promote the formation of medium-sized settlements, particularly for societies with a high degree of technological implementation. Thus, there is a negative correlation between the number of settlements and the size of the largest settlement. 

\section{Conclusions and prospective remarks}
\label{Conclusions}

We have developed an agent-based model to investigate the intricate relationship between technological level and settlement patterns. Through simulations of human societies with varying technological advancements, incorporating collapse events and a saturation term, we have obtained valuable insights into how technological progress and practical constraints shape settlement dynamics. Our findings have revealed the complexity of this relationship, characterized by non-monotonic patterns in the size of the largest settlement and the number of settlements. This model serves as a valuable tool for understanding how technological advancements interact with other factors to influence settlement patterns in human societies.

The results of our study uncovered a significant correlation between the technological level and the shift from pro-family to prosocial behavior. Moreover, the developmental demands imposed by technology exert a profound influence on the nature of this transition, encompassing a spectrum that spans from abrupt to smooth. Importantly, we find that, for enough implementation of the technology, a minimal level of intolerance towards non-prosocial behavior is enough for the prevalence of prosociality regardless of the technological specialization requirements. Also, the results suggest that the connection between technological level and settlement patterns is multifaceted and non-trivial. While higher levels of technological implementation can facilitate the emergence of larger settlements, the specific requirements imposed by the technology can hinder this process. The size of the largest settlement exhibits a local maximum at intermediate technological levels, indicating an optimal point where settlements reach their maximum size. Additionally, the number of settlements initially decreases as the technological level increases, eventually reaching a local minimum. Further to that minimum, as the technological level increases, there is a trend toward a concentration of population in medium-sized settlements. These patterns highlight the presence of a trade-off between technological progress and practical constraints in shaping settlement patterns.

In addition to these findings, several prospective remarks can be made regarding the future direction of research in this area. Firstly, considering non-binary strategies, where individuals can adopt a range of behaviors between pro-family and prosocial, could provide a more nuanced understanding of settlement dynamics. Secondly, incorporating migration into the model would offer insights into how population movements and interactions between different settlements influence settlement patterns and the spread of behaviors. Understanding the interplay between migration patterns and settlement dynamics would provide a deeper insight into the factors shaping human societies.


\section*{Acknowledgments}

\nuevo{We acknowledge partial support from the Government of Aragon, Spain, and “ERDF A way of making Europe” through grant E36-23R (FENOL) and from Ministerio de Ciencia e Innovaci\'on, Agencia Espa\~nola de Investigaci\'on (MCIN/ AEI/10.13039/501100011033) Grant No. PID2020-115800GB-I00 to C.G.L. and Y.M. }

\bibliographystyle{unsrt}

\bibliography{settlements} 

\begin{thebibliography}{10}

\bibitem{parsons1972archaeological}
Jeffrey~R Parsons.
\newblock Archaeological settlement patterns.
\newblock {\em Annual Review of Anthropology}, 1(1):127--150, 1972.

\bibitem{kowalewski2008regional}
Stephen~A Kowalewski.
\newblock Regional settlement pattern studies.
\newblock {\em Journal of Archaeological Research}, 16:225--285, 2008.

\bibitem{ford2009modeling}
Anabel Ford, Keith~C Clarke, and Gary Raines.
\newblock Modeling settlement patterns of the late classic maya civilization
  with bayesian methods and geographic information systems.
\newblock {\em Annals of the Association of American Geographers},
  99(3):496--520, 2009.

\bibitem{burley2015bayesian}
David Burley, Kevan Edinborough, Marshall Weisler, and Jian-xin Zhao.
\newblock Bayesian modeling and chronological precision for polynesian
  settlement of tonga.
\newblock {\em PLoS One}, 10(3):e0120795, 2015.

\bibitem{rodriguez2013modeling}
Guillermo Rodr{\'\i}guez-G{\'o}mez, Jes{\'u}s Rodr{\'\i}guez,
  Jes{\'u}s~{\'A}ngel Mart{\'\i}n-Gonz{\'a}lez, Idoia Goikoetxea, and Ana
  Mateos.
\newblock Modeling trophic resource availability for the first human settlers
  of europe: the case of atapuerca td6.
\newblock {\em Journal of Human Evolution}, 64(6):645--657, 2013.

\bibitem{santos2015effect}
Jos{\'e}~Ignacio Santos, Mar{\'\i}a Pereda, D{\'e}bora Zurro, Myrian
  {\'A}lvarez, Jorge Caro, Jos{\'e}~Manuel Gal{\'a}n, and Ivan Briz~i Godino.
\newblock Effect of resource spatial correlation and hunter-fisher-gatherer
  mobility on social cooperation in tierra del fuego.
\newblock {\em PLoS One}, 10(4):e0121888, 2015.

\bibitem{maher2012pre}
Lisa~A Maher, Tobias Richter, and Jay~T Stock.
\newblock The pre-natufian epipaleolithic: Long-term behavioral trends in the
  levant.
\newblock {\em Evolutionary Anthropology: issues, news, and reviews},
  21(2):69--81, 2012.

\bibitem{pokotylo1978lithic}
David~L Pokotylo.
\newblock {\em Lithic technology and settlement patterns in upper Hat Creek
  Valley, BC}.
\newblock PhD thesis, University of British Columbia, 1978.

\bibitem{boserup2011conditions}
Ester Boserup.
\newblock {\em The conditions of agricultural growth: The economics of agrarin
  change under population pressure}.
\newblock Transaction Publishers, 2011.

\bibitem{bowles2019neolithic}
Samuel Bowles and Jung-Kyoo Choi.
\newblock The neolithic agricultural revolution and the origins of private
  property.
\newblock {\em Journal of Political Economy}, 127(5):2186--2228, 2019.

\bibitem{andrefsky1994raw}
William Andrefsky.
\newblock Raw-material availability and the organization of technology.
\newblock {\em American antiquity}, 59(1):21--34, 1994.

\bibitem{smith2004archaeology}
Michael~E Smith.
\newblock The archaeology of ancient state economies.
\newblock {\em Annu. Rev. Anthropol.}, 33:73--102, 2004.

\bibitem{aldenderfer1993ritual}
Mark Aldenderfer.
\newblock Ritual, hierarchy, and change in foraging societies.
\newblock {\em Journal of Anthropological Archaeology}, 12(1):1--40, 1993.

\bibitem{boyd2011cultural}
Robert Boyd, Peter~J Richerson, and Joseph Henrich.
\newblock The cultural niche: Why social learning is essential for human
  adaptation.
\newblock {\em Proceedings of the National Academy of Sciences},
  108(supplement\_2):10918--10925, 2011.

\bibitem{dandekar2011satvai}
Deepra Dandekar and Abhijit Dandekar.
\newblock The satvai and settlement pattern in rural western maharashtra.
\newblock {\em South Asian Studies}, 27(2):221--224, 2011.

\bibitem{henderson2005muisca}
Hope Henderson and Nicholas Ostler.
\newblock Muisca settlement organization and chiefly authority at suta, valle
  de leyva, colombia: A critical appraisal of native concepts of house for
  studies of complex societies.
\newblock {\em Journal of Anthropological Archaeology}, 24(2):148--178, 2005.

\bibitem{tanimoto2007does}
Jun Tanimoto.
\newblock Does a tag system effectively support emerging cooperation?
\newblock {\em Journal of theoretical biology}, 247(4):756--764, 2007.

\bibitem{pan2013cooperation}
Xiaofei~Sophia Pan and Daniel Houser.
\newblock Cooperation during cultural group formation promotes trust towards
  members of out-groups.
\newblock {\em Proceedings of the Royal Society B: Biological Sciences},
  280(1762):20130606, 2013.

\bibitem{cuesta2015reputation}
Jose~A Cuesta, Carlos Gracia-L{\'a}zaro, Alfredo Ferrer, Yamir Moreno, and
  Angel S{\'a}nchez.
\newblock Reputation drives cooperative behaviour and network formation in
  human groups.
\newblock {\em Scientific reports}, 5(1):7843, 2015.

\bibitem{javarone2017evolutionary}
Marco~Alberto Javarone and Daniele Marinazzo.
\newblock Evolutionary dynamics of group formation.
\newblock {\em PLoS One}, 12(11):e0187960, 2017.

\bibitem{perc2017statistical}
Matja{\v{z}} Perc, Jillian~J Jordan, David~G Rand, Zhen Wang, Stefano
  Boccaletti, and Attila Szolnoki.
\newblock Statistical physics of human cooperation.
\newblock {\em Physics Reports}, 687:1--51, 2017.

\bibitem{javarone2016statistical}
Marco~Alberto Javarone.
\newblock Statistical physics of the spatial prisoner’s dilemma with
  memory-aware agents.
\newblock {\em The European Physical Journal B}, 89:1--6, 2016.

\bibitem{szekely2021evidence}
Aron Szekely, Francesca Lipari, Alberto Antonioni, Mario Paolucci, Angel
  S{\'a}nchez, Luca Tummolini, and Giulia Andrighetto.
\newblock Evidence from a long-term experiment that collective risks change
  social norms and promote cooperation.
\newblock {\em Nature communications}, 12(1):5452, 2021.

\bibitem{szabo2002phase}
Gy{\"o}rgy Szab{\'o} and Christoph Hauert.
\newblock Phase transitions and volunteering in spatial public goods games.
\newblock {\em Physical review letters}, 89(11):118101, 2002.

\bibitem{hauert2008public}
Christoph Hauert, Arne Traulsen, Hannelore De~Silva~n{\'e}e Brandt, Martin~A
  Nowak, and Karl Sigmund.
\newblock Public goods with punishment and abstaining in finite and infinite
  populations.
\newblock {\em Biological theory}, 3:114--122, 2008.

\bibitem{szolnoki2010reward}
Attila Szolnoki and Matjaz Perc.
\newblock Reward and cooperation in the spatial public goods game.
\newblock {\em Europhysics Letters}, 92(3):38003, 2010.

\bibitem{molina2019intergenerational}
Jos{\'e}~Alberto Molina, Alfredo Ferrer, J~Ignacio Gim{\'e}nez-Nadal, Carlos
  Gracia-L{\'a}zaro, Yamir Moreno, and Angel Sanchez.
\newblock Intergenerational cooperation within the household: A public good
  game with three generations.
\newblock {\em Review of Economics of the Household}, 17(2):535--552, 2019.

\bibitem{perc2013evolutionary}
Matja{\v{z}} Perc, Jes{\'u}s G{\'o}mez-Gardenes, Attila Szolnoki, Luis~M
  Flor{\'\i}a, and Yamir Moreno.
\newblock Evolutionary dynamics of group interactions on structured
  populations: a review.
\newblock {\em Journal of the royal society interface}, 10(80):20120997, 2013.

\bibitem{bowles2006group}
Samuel Bowles.
\newblock Group competition, reproductive leveling, and the evolution of human
  altruism.
\newblock {\em science}, 314(5805):1569--1572, 2006.

\bibitem{axelrod2006agent}
Robert Axelrod.
\newblock Agent-based modeling as a bridge between disciplines.
\newblock {\em Handbook of computational economics}, 2:1565--1584, 2006.

\bibitem{epstein2006remarks}
Joshua~M Epstein.
\newblock Remarks on the foundations of agent-based generative social science.
\newblock {\em Handbook of computational economics}, 2:1585--1604, 2006.

\bibitem{schelling1971dynamic}
Thomas~C Schelling.
\newblock Dynamic models of segregation.
\newblock {\em Journal of mathematical sociology}, 1(2):143--186, 1971.

\bibitem{axelrod1997dissemination}
Robert Axelrod.
\newblock The dissemination of culture: A model with local convergence and
  global polarization.
\newblock {\em Journal of conflict resolution}, 41(2):203--226, 1997.

\bibitem{parker2003multi}
Dawn~C Parker, Steven~M Manson, Marco~A Janssen, Matthew~J Hoffmann, and Peter
  Deadman.
\newblock Multi-agent systems for the simulation of land-use and land-cover
  change: a review.
\newblock {\em Annals of the association of American Geographers},
  93(2):314--337, 2003.

\bibitem{gracia2022dynamics}
Carlos Gracia-L{\'a}zaro, Fabio Dercole, and Yamir Moreno.
\newblock Dynamics of economic unions: An agent-based model to investigate the
  economic and social drivers of withdrawals.
\newblock {\em Chaos, Solitons \& Fractals}, 160:112223, 2022.

\bibitem{johnson2000evolution}
Allen~W Johnson and Timothy~K Earle.
\newblock {\em The evolution of human societies: from foraging group to
  agrarian state}.
\newblock Stanford University Press, 2000.

\bibitem{ellis2015ecology}
Erle~C Ellis.
\newblock Ecology in an anthropogenic biosphere.
\newblock {\em Ecological Monographs}, 85(3):287--331, 2015.

\bibitem{verhulst1845resherches}
Pierre~Fran{\c{c}}ois Verhulst.
\newblock Resherches mathematiques sur la loi d'accroissement de la population.
\newblock {\em Nouveaux memoires de l'academie royale des sciences}, 18:1--41,
  1845.

\bibitem{del2004carrying}
Pablo Del Monte-Luna, Barry~W Brook, Manuel~J Zetina-Rej{\'o}n, and Victor~H
  Cruz-Escalona.
\newblock The carrying capacity of ecosystems.
\newblock {\em Global ecology and biogeography}, 13(6):485--495, 2004.

\end{thebibliography}

\renewcommand{\thefigure}{SI-\arabic{figure}}
\setcounter{figure}{0} 

\linespread{1.9}

\newpage

{\huge
\begin{center}
    
\textbf{Supplementary Information}
\end{center}

}

\section*{I Sensitivity Analysis with Larger Population Size}
\vspace{3mm}

In the main text, we presented simulations with a population size of 10,000 agents. In this section, we present the results of simulations with a larger population size of 100,000 agents, explicitly focusing on the stationary prosociality as a function of the technological level $\alpha$ for two different values of tolerance: $T=1$ and $T=0.5$.

\begin{figure}[h]
\centering
    \includegraphics[width=0.48\textwidth]{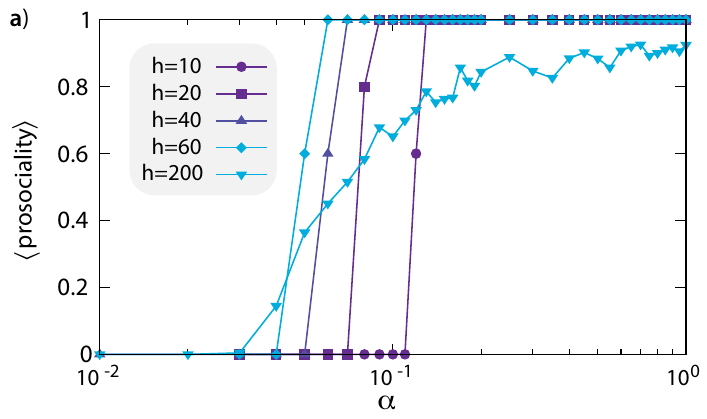}
    \includegraphics[width=0.48\textwidth]{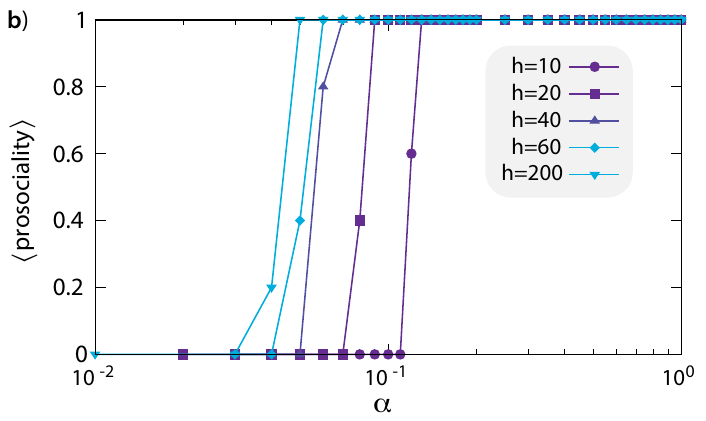}
    \caption{Fraction of prosocial agents versus the technological level $\alpha$ for a population size of $n=10^5$ agents. Panels \textbf{a} and \textbf{b} correspond to tolerance values of $T=1$ and $T=0.5$, respectively. The curves represent the fraction of prosocial agents at equilibrium as a function of $\alpha$, with different colors representing different values of the saturation parameter $h$. The simulations start with 50\%-50\% of prosocial and pro-family agents.}
\label{fig:SI_c_Vs_alpha_n100000}
\end{figure}

Figure \ref{fig:SI_c_Vs_alpha_n100000} shows the results of these simulations, with panels \textbf{a} and \textbf{b} corresponding to $T=1$ and $T=0.5$, respectively. As in the main text, the curves represent the fraction of prosocial agents at equilibrium as a function of $\alpha$, with different colors representing different values of the saturation parameter $h$.

The results show that the behavior of prosociality versus $\alpha$ is qualitatively equivalent to the results in the main text. As $\alpha$ increases, there is a transition from prosocial to pro-family behavior, with the critical value of $\alpha$ decreasing as $h$ increases. For $T=1$, the transition is smooth only for high values of $h$, whereas, for $T=0.5$, the transition is abrupt for all values of $h$. 
Despite the overall similarities between Figure \ref{fig:SI_c_Vs_alpha_n100000} and Figure 2 of the main text, a closer comparison suggests that the curves for high $h$ without collapses ($T=1$) depend on population size. Near the critical point, larger sizes may increase the probability for a settlement to have the conditions for prosociality to thrive, resulting in more abrupt transitions for high values of $h$. This is supported by the probabilities for early families of different strategist configurations (as shown in Section \ref{section:binomial_destribution}), which involve a higher likelihood for prosocial behavior to emerge in larger settlements. From a physics point of view, that fact could indicate that the nature of the transition in the thermodynamic limit could be of first order even for large $h$ and $T$ values. However, it is important to keep in mind that in practical applications of the model, finite-size effects will always be present.

\newpage
\section*{II Sensitivity Analysis with Different Initial Conditions}

\vspace{3mm}

In addition to the simulations with initial equiprobable strategies presented in the main text, we conducted additional simulations in which only 5\% of agents initially adopted prosocial behavior and 95\% adopted pro-family behavior. These simulations allow us to study the fixation of prosociality in a realistic scenario where prosocial behavior is initially rare, resembling the formation of settlements only when favorable conditions are present.

\begin{figure}[h]
  \centering
    \includegraphics[width=0.48\textwidth]{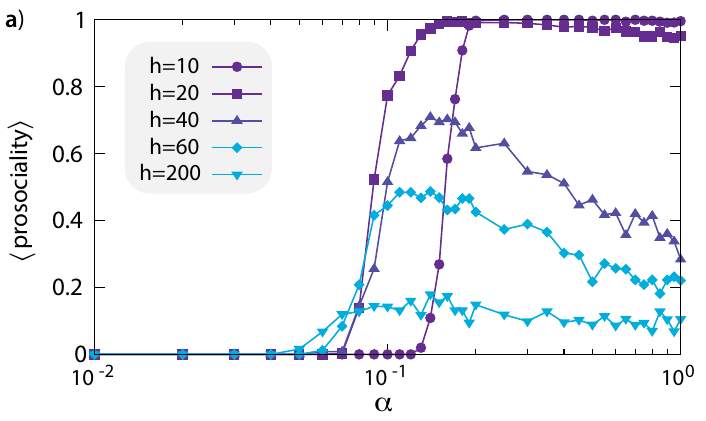}
    \includegraphics[width=0.48\textwidth]{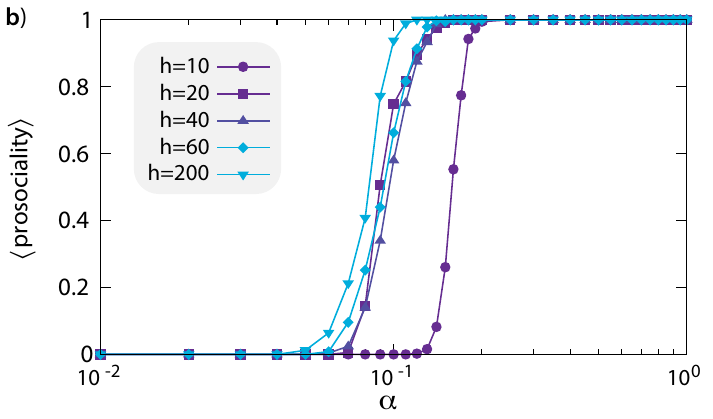}
    \includegraphics[width=0.48\textwidth]{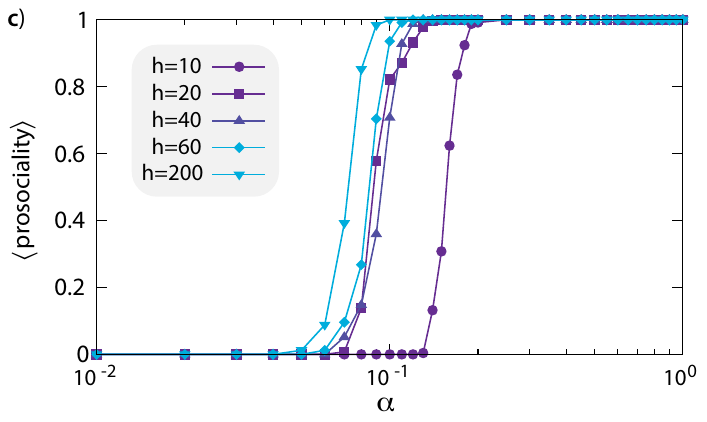}
    \includegraphics[width=0.48\textwidth]{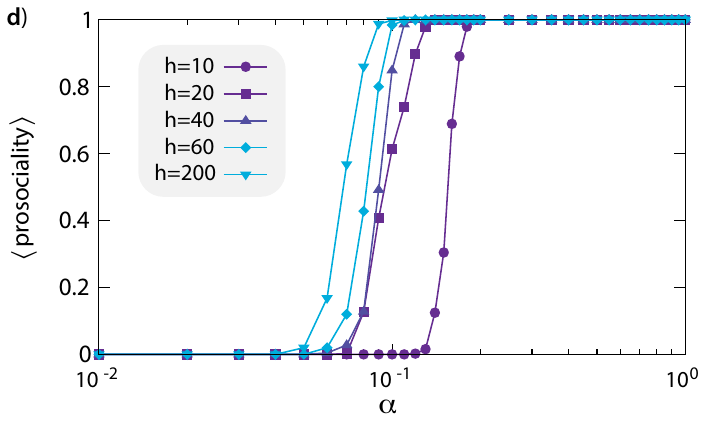}
    \caption{Fraction of prosocial agents as a function of the technological level $\alpha$ for a low initial prosociality. Panels \textbf{a}, \textbf{b}, \textbf{c}, and \textbf{d} display the results for tolerance values of $T=1,0.95, 0.9, 0.5$, respectively. Each series corresponds to a different saturation value $h=10, 20, 40, 60, 80, 200$, and each point represents the mean value over 100 numerical simulations. The initial condition is 5\% of prosocial and 95\% of pro-family agents. These plots correspond to those in Figure 2 in the main text, where the initial distribution of strategies was 50\%-50\% and the tolerance values were $T=1,0.95, 0.9, 0.5$. The total number of agents in the simulations is $n=10^4$.}
  \label{fig:SI_c_Vs_alpha}
\end{figure}

Figure \ref{fig:SI_c_Vs_alpha} corresponds to Figure 2 in the main text, but with an initial condition of 5\% prosocial agents. It displays the final prosocial ratio as a function of the technological level $\alpha$. Each series corresponds to a different saturation value $h$ ranging from 10 to 200, and each point represents the mean value over 100 numerical simulations with a population size of $n=10^4$. Panels \textbf{a}, \textbf{b}, \textbf{c}, and \textbf{d} correspond to tolerance values $T$ of 1, 0.95, 0.9, and 0.5, respectively. 
Before analyzing the curves, let us note that an initial frequency of only 0.05\% prosocial agents implies that i) the probability of initially having at least one all-prosocial family is very low ($p\sim 6.3 \times 10^{-4}$), and ii) it is highly likely ($p\sim 1$) that some families initially will only have a pro-family agent. Additionally, the initially expected fraction of fully-pro-family families is 77\%
. See Section \ref{section:binomial_destribution} for details.

The main differences between Figure \ref{fig:SI_c_Vs_alpha} and Figure 2 in the main text are observed in panel \textbf{a} ($T=1$), which represents scenarios without collapses. For $T=1$ and intermediate to large values of $h$, it is not guaranteed that the society will eventually go to a fully prosocial state, even for high values of $\alpha$. As discussed earlier, in the initial stages, there may be some families with high prosociality, and fluctuations can eventually lead them toward a fully prosocial configuration. However, in the absence of initially fully prosocial families, the characteristic times associated with this process will be longer compared to scenarios with an equiprobable initial distribution of strategies, where fully prosocial families are initially present, as shown in Section \ref{section:binomial_destribution}. Consequently, in the absence of collapses, especially when a large settlement size is required for prosocial behavior to reach its maximum performance (intermediate and large $h$ values), the likelihood of these prosocial settlements dominating the system before achieving full prosociality will be smaller compared to scenarios with an equiprobable strategies distribution. Instead, there is a higher probability of dominating the system without reaching full prosociality. As discussed in the main text, in such scenarios, prosociality will gradually decline toward zero.

However, when collapses are introduced ($T<1$, panels \textbf{b}-\textbf{d}), once the prosociality of that dominant settlement decreases to the value of $1-T$, the settlement collapses. After the collapse, with all families becoming mono-strategic, some families will be fully prosocial, forming the first prosocial settlements that possess a significant evolutionary advantage over exclusively pro-familiar settlements. This substantial difference in fitness will enable the largest prosocial settlement to grow rapidly and ultimately dominate the system.

Furthermore, regarding $T<1$ (panels \textbf{b}-\textbf{d}), the similarity of curves in Figure \ref{fig:SI_c_Vs_alpha}, where the probability of initially having at least one fully-prosocial family is negligible, with those in Figure 1 of the main text, where the probability of initially having at least one fully-prosocial family is very high ($p\sim 1$), suggests that fully-prosocial families do not play a significant role in driving the dynamics towards prosocial behavior in the initial stages.


\section*{III Probabilistic analysis of families' strategies: impact of\\initial conditions}
\label{section:binomial_destribution}
\vspace{3mm}

For the case of 95\% of pro-family agents, 5\% of prosocial agents, and a population of $n=10^4$ agents, in the initial state ($t=0$):

\begin{enumerate}
\item The probability $p_0$ for a given 5-member family having no pro-family members is $p_0=0.05^5\sim 3 \times 10^{-7}$. Therefore, the probability of having at least a pro-family member is $1-p_0 \sim 1$. 

Let $a$ be the number of families, initially $a=2000$. The probability for all families having at least a pro-family member is $(1-p_0)^a\sim0.999374$,
and the probability of being at least a fully-prosocial family is given by:

$P(X \geq 1) = 1-(1-p_0)^a \sim 6.3 \times 10^{-4}$\;

Therefore, the probability of having at least one family without any pro-family members is very low: 0.063\%.

\item For a given 5-member family, the probability $p_{1}$ of having one (and only one) pro-family member is given by a binomial ($p=0.95, q=0.05$) distribution with 5 trials:

$p_{1} = {5 \choose 1}\;p^4\;q^1= 5\cdot 0.95^4 \cdot 0.05 = 0.2375 $\;.

The probability $p_{0,1}$ for a given family having at least one pro-family member is given by:

$p_{0,1} = p_{0} + p_{1} \sim p_{1} $\;.

The likelihood of there being at least one family with either one or zero pro-family members among $a = 2000$ families is determined by:

$1-(1-p_{0,1})^a \sim 1 $\;.

Therefore, the probability of having at least one family with one or no pro-family members is approximately 1.

\item Let $V$ be the probability of a given family being fully pro-family, and $U$ be the expected number of families with no prosocial members.

$V = (0.95)^5 \cdot (0.05)^0 \sim 0.77$\;.

$U = a \cdot V= 2000 \cdot  V \sim 1548$\;.

So the expected fraction of families without prosocial members is approximately 0.77, i.e., 77\% of the families.

\end{enumerate}

\noindent
The same calculations can be applied to the case of the equiprobable initial distribution of strategies (50\%-50\% of pro-family and prosocial agents) and population size $n=10^4$. In this later scenario:

\begin{enumerate}
\item The probability of having at least one family without any pro-family member is approximately 100\%
\item The probability of having at least one family with one or no pro-family members is approximately 100\%
\item The expected value of families without prosocial agents is 62.5 (i.e., 3.125\% of the families).
\end{enumerate}

\end{document}